\documentclass[a4paper,12pt]{article}
\usepackage{graphicx}%
\usepackage{amsmath}
\begin{document}
\begin{center}
{\bf \large  Study of two photon production process in proton-proton
collisions at 216 MeV}
\vskip 5mm
{\large \bf A.S.Khrykin}\\[0.25 cm]
{ \small \it Joint Institute for Nuclear Research, Dubna, 141980
Russia\\
Email: Khrykin@nusun.jinr.dubna.su}
\end{center}
\begin{abstract}
The energy spectrum for high
energy $\gamma$-rays ($E_\gamma \geq 10$ MeV) from the process $pp
\to \gamma \gamma X$ emitted at $90^0$ in the laboratory frame
has been measured at 216 MeV. The resulting photon energy
spectrum extracted from $\gamma-\gamma$ coincidence events
consists of a narrow peak (5.3$\sigma$) at a photon energy of
about 24 MeV and a relatively broad peak (3.5$\sigma$) in the energy range of
(50 - 70) MeV.
This behavior of the photon energy
spectrum is interpreted as a signature of the exotic dibaryon
resonance $d^\star_1$ with a mass of about 1956 MeV which is assumed
to be formed in the radiative
process $pp \to \gamma \ d^\star_1$ \ followed by its electromagnetic decay
via the
$d^\star_1 \to pp \gamma$ \ mode.
The experimental spectrum is compared with those obtained by means
of Monte Carlo simulations.
\end{abstract}
\section{Introduction}
The process $pp \to pp \gamma\gamma$
at energy below the pion production threshold ($\pi NN$)
 is still poorly explored both theoretically and experimentally.
It was first suggested not long ago as a sensitive probe of
the possible existence of $NN$-decoupled nonstrange dibaryon
resonances\cite{GerKh93,EGK95}. These are two-baryon
states $^2B$ with zero strangeness and exotic quantum numbers
$I(J^P)$ [$I$ is the isospin, $J$ is the total spin, and $P$ is
the parity of a dibaryon state], for which the strong decay
$^2B \to pp$ \ is either strictly forbidden by the Pauli
principle [for the states with $I(J^P)=1(1^+,3^+,etc.$)]
or is strongly
suppressed by the isospin selection rules (for the states with
$I=2$). Such dibaryon
states cannot be simply bound systems
of two nucleons, and
a proof of their existence would have consequences of fundamental
significance for the theory of strong interactions\cite{QCDinm,Kondr,Kopel}.

If the $NN$-decoupled dibaryons
exist in nature, then the $pp\gamma\gamma$ \ process may proceed, at
least partly, through the mechanism that directly involves the
radiative excitation $pp \to \gamma \ ^2B$ \ and decay $^2B \to
\gamma pp$ \ modes of these states.
In $pp$ collisions at energies below the $\pi NN$ \
threshold, these production and decay modes of
the $NN$-decoupled dibaryon resonances with masses
$M_R \leq 2m_p + m_\pi$ \ would be unique or dominant.
Since such dibaryons may decay mainly into the $pp\gamma$ state,
their widths should be very narrow ($\le 1 keV$).
The simplest and clear way of revealing them
is to measure the photon energy spectrum of the
reaction $pp\gamma\gamma$. The presence of an $NN$-decoupled dibaryon
resonance would reveal itself in this energy spectrum as a narrow peak associated with
the formation of the resonance and a relatively
broad peak originating from its three-particle decay.
In the center-of-mass system, the position of the narrow peak ($E_R$)
is determined by the energy of colliding nucleons ($W=\sqrt{s}$)
and the mass of this dibaryon resonance as $E_R=(W^2-M_R^2)/2W$.
An essential feature of the $pp\gamma\gamma$ \ process
at an energy below the $\pi NN$ \  threshold is that,
apart from the resonant mechanism in question, there should only
be one more source of photon pairs.
This is the double $pp$ \  bremsstrahlung reaction.
But this reaction is expected to play a minor role.
Indeed, it involves two electromagnetic vertices, so that one may
expect that the $pp\gamma\gamma$-to-$pp\gamma$ \ cross section ratio
should be of the order of the fine structure constant $\alpha$.
However, the cross section for $pp\gamma$ \ is already
small (the total cross section for the $pp\gamma$ \ reaction at energies
of interest is a few $\mu b$).

The preliminary experimental studies of the
reaction $pp \to \gamma\gamma X$ \  at an
energy of about 200 MeV \cite{Panic96,Menu97} showed that the photon energy
spectrum of this reaction had a peculiar structure ranging from
about 20 MeV to about 60 MeV. This structure was interpreted as
an indication of the possible existence of an $NN$-decoupled dibaryon
resonance (later called $d^\star_1$) that is produced
in the process $pp \to \gamma \ d^\star_1$ \ and subsequently decays via
the $d^\star_1 \to pp \gamma$ \ channel.
Unfortunately, a relatively
coarse energy resolution and low statistics did not allow
us to distinguish the narrow $\gamma$ peak associated with
the $d^\star_1$ production from the broad $\gamma$ peak due to its decay
 and, hence, to determine the resonance mass exactly.
To clarify the situation with the dibaryon
resonance $d^\star_1$, we have decided to measure the energy spectrum
of the $pp \to pp\gamma\gamma$ \ reaction more carefully.
\section{The Experiment and Results}
The experiment was performed using the variable energy proton
beam from the phasotron at the Joint Institute for Nuclear
Research (JINR).
 The pulsed proton beam with an energy of about 216 MeV,
an energy spread of about 1.5\%, and an intensity of about
$3.6 \cdot 10^8$ protons/s bombarded a liquid hydrogen target.
Both $\gamma$ quanta of the reaction $pp\gamma\gamma$ were detected by
two $\gamma$-ray detectors placed
in a horizontal plane, symmetrically on either side of the beam
at a laboratory angle of $90^0$ with respect
to the beam direction.
 The solid angles covered by
the detectors were 43 msr and 76 msr, respectively.
To reject events induced by charged particles, plastic scintillators
were put in front of each $\gamma$ detector.
The electronics associated with the
$\gamma$ detectors and the plastic scintillators
together with the data acquisition system provided
$\gamma - \gamma$ \ coincidence candidate events to be recorded on
the hard disk of the
computer. A further selection of events associated with the
process $pp \to \gamma\gamma X$ \ was done during off-line data
processing. The energy threshold for both the
$\gamma$ detectors was set at about 7 MeV.

Measurements were done both for the target filled with liquid hydrogen
and for the empty one.
Data for the full and empty target were taken
in two successive runs for $\sim$31 h and $\sim$21 h, respectively.
The integrated luminosity of about $8.5\ pb^{-1}$ was
accumulated for the measurement with the full target.
The photon energy spectrum of the process $pp \to \gamma \gamma X$
obtained after subtraction of the empty-target
contribution from the spectrum measured with the full target
is shown in Fig. 1.
As can be seen, this spectrum
consists of a narrow peak at a photon energy of about 24
MeV and a relatively broad peak in the energy range from about 50 to
about 70 MeV. The statistical significances
for the narrow and the broad peaks are 5.3$\sigma$ and 3.5$\sigma$,
respectively.
The width (FWHM) of the narrow peak
was found to be about 8 MeV. This width is comparable with that
of the energy resolution of the experimental setup.
The observed behavior of the photon energy
spectrum agrees with a characteristic signature of the
sought dibaryon resonance $d^\star_1$ that is formed and decays
in the radiative process $pp \to \gamma \ d^\star_1 \to pp \gamma\gamma$.
In that case the narrow
peak should be attributed to the formation of
this dibaryon, while the broad peak should be assigned
to its three-particle decay. Using
the value for the energy of the narrow peak $E_R \sim 24$ MeV, we obtained
the $d^\star_1$ mass  $M_R \sim 1956$ MeV.
The differential
cross section for the resonance production of two photons
emitted symmetrically at $\theta_{lab} = \pm 90^0$ from
the process $pp \to \gamma \ d^\star_1 \to pp \gamma\gamma$
at an energy of 216 MeV was estimated to be $\sim$ 9 nb/sr$^2$.

Having assumed that the $pp \to \gamma \ d^\star_1 \to pp
\gamma\gamma$ \ process with the $d^\star_1$ mass of 1956 MeV is the only
mechanism of the reaction $pp \to pp \gamma\gamma$, we calculated
the photon energy spectra of this reaction for a proton energy of
216 MeV. It was also assumed that the
radiative decay of the $d^\star_1$ is a dipole $E1 (M1)$ transition
from the two-baryon resonance state to a $pp$ state in the continuum.
The calculations were carried out with the help of Monte Carlo
simulations which included the geometry and the energy resolution of
the actual experimental setup. The photon energy spectra were
calculated for two different scenarios of the $d^\star_1$ decay.
The difference between them was that one of these scenarios took into
account the final state interaction ($FSI$) of two outgoing
protons whereas in the other that interaction was switched off.
Each of the
scenarios imposed some restrictions on possible quantum numbers of
the dibaryon state in question. The scenario including the $FSI$
implies that the final $pp$-system is in the singlet $^1S_0$
state and consequently it should take place, in particular,
for the isovector
$1^+$ dibaryon state (the simplest exotic quantum numbers), namely,
$1^+ \stackrel{M1}{\longrightarrow} 0^+$. Moreover such a scenario
can take place for any isotensor dibaryon state with the exception
of the $0^+$ or $0^-$ state.
 At the same time, the scenario
in which the $FSI$ is switched off, is most likely to occur for
the isotensor $0^\pm$ dibaryon state.
The spectra calculated for these two decay scenarios
 and normalized to the total number of $\gamma-\gamma$
events observed in the present experiment are shown in Fig. 1.
\begin{figure}[t]
\begin{center}
\includegraphics[width=80 mm]{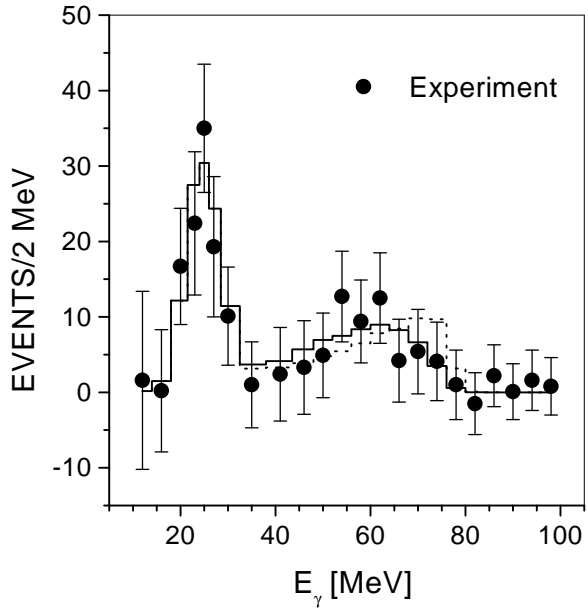}%
\end{center}
\caption{\small{Experimentally observed
energy spectrum for photons from the $pp\gamma\gamma$ \
process and energy spectra for photons from the process
$pp \to \gamma d^\star_1 \to \gamma\gamma pp$ \
calculated with the help of Monte Carlo simulations for two $d^\star_1$
decay scenarios: without the FSI (solid line)
and with the FSI (dashed line).}}
\end{figure}
Comparison of these spectra with the experimental spectrum indicates
that both the calculated spectra are in reasonable agreement with
the experimental one within experimental uncertainties.
In other words, the statistics of the
experiment is insufficient to draw any firm conclusions in favor
of one of these scenarios and thereby to limit possible
quantum numbers of the observed dibaryon state.

Here it is important to note that the $d^\star_1$
with any possible set of quantum numbers $I(J^P)$
with the exception of $2(0^\pm)$
should mainly decay to the singlet $^1S_0$ pp state. At the same
time the outgoing protons from the process $pp \to \gamma \ d^\star_1 \to \gamma\gamma ^1S_0{pp}$
would mainly be concentrated in a
narrow angular cone near the direction of motion of incident
protons \cite{PRC64}. We believe that this is why that this process was not
found in the Uppsala $pp$ bremsstrahlung data \cite{WASA}.
\section{Conclusion}
 The $\gamma$-ray energy spectrum for the $pp \to \gamma \gamma X$ \ reaction
at a proton energy below the pion production threshold has been measured
for the first time.
The spectrum measured at an energy of about 216 MeV
for coincident photons emitted at an angle of $90^0$ in the laboratory frame
clearly evidences the existence of the $NN$-decoupled dibaryon
resonance $d^\star_1$ with a mass of $\sim 1956$ MeV that is
formed and decays in the process
$pp \to \gamma \ d^\star_1 \to pp \gamma\gamma$.
The data we have obtained, however, are still incomplete, and
additional careful studies of the reaction
$pp \to pp \gamma \gamma$ \ are needed to get proper
parameters (mass, width, spin, etc.) of the observed dibaryon state.

\end{document}